# Indicators of the human origin of numbers

Vitaliy Grigoriev, Department of Sociology, St.Petersburg State University

v.grigorev@spbu.ru

**Abstract**

Researchers have demonstrated that humans are unable to generate a sequence of random numbers that corresponds in a statistical sense to a simple distribution such as the uniform distribution. The purpose of this article is to present the results of research on the generation of random number sequences by humans. The article describes 10 effects found in such studies, mechanisms explaining these effects, and 14 measures (not including modifications) used to detect deviations from randomness in the sequences. The analysis of numerical sequences is not only of academic interest; it can also be used for the purpose of data validation (auditing).

Researchers agree that people are not able to create sequences of numbers that would have the property of randomness. The nature of the deviations of sequences obtained from humans in the process of random number generation (RNG) from a uniform distribution is interesting in several ways. First, the study of deviations sheds light on the cognitive mechanisms of the human brain (Heuer et al., 2005; Sexton, Cooper, 2014; Cai, Li, 2015). There are three mechanisms that affect RNG: (1) the *inhibition mechanism* responsible for the suppression of stereotypical sequences, (e.g. 1, 2, 3, 4); (2) a *memory updating mechanism* needed to control the representation of numbers in a sequence; (3) the *shifting mechanism* responsible for mental movement between objects in sequences (Towse and Neil, 1998; Miyake, et al., 2000).

Researchers suggest that the level of randomness of sequences may be influenced by functional limitations of subjects: insufficient memory volume (Tune, 1964; Towse, Cheshire, 2007) or inability to maintain attention (Weiss, 1964). A small volume of memory increases the chance of stereotypical responses, but poor memory can hinder the tendency to avoid repetitions and actually improve the quality of a random sequence, which was demonstrated when hypnosis was applied to subjects (Terhune, Brugger, 2011). Mental patients are usually worse at coping with RNG than healthy controls. Thus, in early schizophrenia, the stereotypical responses increase, while the ability to recognize the frequency of digits remains intact (Chan et al., 2011). Patients with autism are more likely than healthy people to repeat and cycle (using all alternatives as soon as possible) (Williams et al., 2002). But there are exceptions.

Patients with speech disorder (aphasia) showed better results in generating random sequences in terms of cyclicity compared with healthy controls (Proios et al., 2008).

Second, RNG task is related to the problem of identifying random sequences, which means it is related to how a person perceives events (random or systematic). This perception influences decision-making and behavior. The study of deviations from randomness sheds light on the conceptual understanding of randomness by humans (Kahneman, Tversky, 1972; Diener, Thompson, 1985; Ladouceur, et al., 1996; McDonald, Newell, 2009).

Third, - knowledge of the nature of typical deviations is used in the construction of classifiers that divide sequences into "natural" and "artificial". Such classifiers are used in audit tasks, search for falsified data and other similar tasks (Evans, 1996; Bolton, Hand, 2002; Beber, Scacco, 2008).

The result of the RNG depends on many factors (Wagenaar, 1972):

- the number and type of alternatives (allowed options);

- the length of the required sequence;

- the formulation of the task;

- the generation rate, as well as the individual characteristics of the subject (Chapanis, 1953; Evans, 1978; Towse et al., 2014; Cai, Li, 2015), including ethnicity (Strenge et al., 2009; Wagner, Jamsawang, 2011).

Therefore a greater number of measures have been developed to identify deviations of a different origins.

Next, we will list the main deviations of sequences created by humans from truly random sequences recorded by researchers.

**Characteristics of sequences created by humans**

*Preferences in digits*. Studies show that people prefer certain digits (Chapanis, 1953; Baddeley et al., 1998). For example, when asked to name the "first digit that came to mind", 28.4% of respondents chose **7** and only 2.2% chose **1** (Kubovy, 1977). Researchers note the special status of zero (low popularity) (Boland, Hutchinson, 2000).

*Left bias*. Humans systematically prefer small numbers to large and zero, as well as small digits in numbers to large digits (Rata, 1966; Hill, 1988; Boland, Hutchinson, 2000; Loetscher, Brugger, 2007; Towse et al., 2014).

*Anchoring*. When people were asked to name the first number that came to mind, the number 7 dominated - 28.4% of the responses (Kubovy, Psotka, 1976). If the experimenters mentioned 7, as an example of a possible answer, its frequency decreased (to 16.6%). After 7-tip, it didn't look like a spontaneous choice. If the binding was introduced surreptitiously, then the effect could be the opposite. Asking to name the "first one-digit number that came to mind" increased the proportion of ones from 2.2% to 18.0% compared to asking to name the "first digit that came to mind" (Kubovy, 1977). The phrase ""first one-digit number" made 1 more accessible.

*Avoiding repetitions*. In problems of generating random numbers, the frequency of repetitions (such as: 2, 2) is regularly lower than follows from probability theory (Chapanis, 1953; Rath, 1966; Noether, 1987; Brugger et al., 1996; Towse, 1998; Boland and Hutchinson, 2000). For example, in a study by Philip Boland and Kevin Hutchison, 70% of respondents did not repeat a single digit in a row in a 25-digit sequence. Although there should be about 8% of such random sequences. In addition, multi-digit numbers show a tendency to use different digits (e.g. 752, 312, 587).

*Avoiding long monotonous sequences*. In sequences created by humans, the frequency of switching, that is, a change in the direction of increasing sequence to decreasing and vice versa, is more common than should be expected according to probability theory (Falk, Konold, 1997).

*Avoiding symmetrical patterns*. Humans avoid symmetric sequences (e.g. 1,3,5,3,1) (Wagenaar, 1972).

*Cyclicity*. The phenomenon is that the subjects tend to use all available alternatives as soon as possible during generation. For example, if the choice is limited to numbers from 1 to 6, the subjects use all the options in the first 6 attempts. The phenomenon expresses the desire for "balance". In one experiment (Borland and Hutchinson, 2000), it was found that 49% of students in a sequence of 25 digits of 12 or 13 digits (closest to half) were located below the theoretical median. While the theory predicts that there should be about 31% of such cases.

*The use of adjacent digits*. In Gustav Rath's experiment, subjects used adjacent numbers more often than expected (12, 23, and the like) (Rath, 1966). In human-generated numbers, decreasing sequences of digits (such as 987, 876, 765, 654) are more popular than increasing ones (such as 456, 567, 678) (Chapanis, 1953).

*Stereotyping*. Subjects tend to reproduce certain combinations. Especially at long intervals, when there are no memories of a certain sequence in memory, it can occur over and over again. In the experiment of Alphonse Chapanis, autocorrelations of numbers up to the 4th order were observed (Chapanis, 1953).

The deviations in question are caused by various reasons and are quite diverse. Researchers solve different problems when working with numerical sequences. Therefore, it is not possible to recommend the best indicator characterizing the differences between human numerical sequences and random ones. We will look at the most widely used multidirectional indicators.

**Indicators of deviations from randomness**

Several reviews of indicators used to analyse deviations from randomness have been published (Towse, Neil, 1998; Mateus, Caeiro, 2014). This review builds on these reviews, taking into account the use of indicators in empirical research in the social sciences.

## *Redundancy (R)*

In information theory, a sequence of items contains maximum first-order information if each option is chosen with equal probability. The amount of first-order information is calculated as:

$$H_{single} = \log_2 n - \frac{1}{n}\left(\sum n_i \log_2 n_i\right),$$

where n is the number of random answers in the set, $n_i$ is the frequency of occurrence of the i-th alternative. The maximum possible amount of information (*a* is the number of possible alternatives),

$$H_{max} = \log_2 a .$$

Accordingly, the redundancy (R) in the number sequence is defined as

$$R = 100 \times \left(1 - \frac{H_{single}}{H_{max}}\right).$$

Redundancy is zero if the sequence contains the maximum possible amount of information. And it increases with decreasing diversity of sequence elements.

## *Coupon*

This indicator reflects the number of numbers generated until the alternatives are completely exhausted. If an alternative does not appear at all, the indicator is equal to the length of the sequence.

## *Random Number Generation Index (RNG)*

RNG (Evans, 1978) is based on measuring how often one alternative follows another. A pair of alternatives is called a digram

$$RNG = \frac{\sum n_{ij} \log n_{ij}}{\sum n_i \log n_i}$$

$n_{ij}$ - shows how often alternative j follows alternative i; $n_i$ - frequency of alternative i. Only non-zero frequencies are included in the calculation. The RNG ranges from 0 (uniformity of digrams) to 1 (complete predictability of pairs).

There is also a Null-Score Quotient (NSQ) based on the number of digrams missing from the series generated by the subject. Obviously, the NSQ is closely related to the RNG and provides similar information. Its disadvantage is that it does not work for very long and very short sequences.

## *Adjacency (A)*

This measure is sometimes referred to as the "stereotypy score"

$$A = 100 \times \frac{number\ of\ adjacent\ pairs}{number\ of\ response\ pairs} .$$

A pair is two adjacent numbers in a sequence. Adjacent pairs are pairs whose values differ by ±1.

### *Count score (CS1, CS2)*

These indices are related to adjacency in terms of the deviation found. CS1 and CS2 are the sums of the squares of the lengths of subsequences containing elements that differ by 1 and 2, respectively. The squaring gives relatively more weight to long sequences than to short sequences (Spatt, Goldenberg, 1993).

### *Turning Point Index (TPI)*

It is based on the number of changes of direction of a numerical series from ascending to descending and vice versa. The sequence "1, 3, 5, 7, 8, 6" contains a single change of direction at the answer "8", from which the descending series starts. The sequence "5, 3, 4, 6, 2, 8, 9, 7" contains 4 turns (at "3", "6", "2" and "9"). In case of repetition, one turn is "extended": (e.g. "2, 4, 4, 4, 3" contains one turn between two "4s").

The TPI value is expressed as a percentage, indicating the correlation between the anticipated and observed number of turns

$TPI = 100 \times \frac{TP_{observed}}{TP_{expected}}$, expected theoretical value - $TP_{expected} = \frac{2}{3}(n-2)$, where n is the length of the sequence. Deviations from 100 may indicate a generation strategy. A TPI less than 100 may indicate a counting strategy, while a TPI greater than 100 may indicate a strategy of alternating small and large numbers.

### *Sign Difference (SIGN)*

Count the number of values in the sequence for which $X_i$ is greater than $X_{i-1}$ or vice versa. This method can detect the same trends as the Turning Point Index.

### *Cox-Stewart Statistic (CSS)*

The data are grouped into pairs, where the i-th observation from the first half of the data is paired with the i-th observation from the second half of the data, sorted by order of generation. In vectors with odd lengths, the middle value is excluded. This statistic is the sign difference applied to such pairs, allowing for the identification of macro trends in number generation.

### *Phase Length (PL)*

The distance between the turning points is referred to as the 'phase'. Consider the sequence '2, 3, 5, 4, 5, 6, 7, 8, 6, 1, 3'. The turning points are values 5, 4, 8, and 1. The phase lengths (PL) between these points are 1, 4, and 2. The PL distribution is calculated for the complete sequence, including the unknown phase lengths at the beginning and end. The expected frequency of phases with length d is

$frequency(d)_{expected} = \frac{2(n-d-2)(d^2+3d+1)}{(d+3)!}$

Next, the expected and empirical distribution of phase lengths are compared, for instance, by using Kolmogorov's criterion of agreement. The average phase lengths of a sequence are related to TPI.

## *Runs*

The measure reflects the diversity of lengths of ascending subsequences. It determines the number of elements in consecutive ascending subsequences. If a number is not included in an ascending subsequence, it is considered a rise of unit length. The index of the rises is the variance of these numbers. This index can be maximized if all variants are placed in strictly increasing order (which is not always possible), and minimized (0) if they are in strictly decreasing order.

## *Wald–Wolfowitz Runs Test (RWW)*

The given data is transformed into a dichotomous vector based on its position relative to the threshold. Any data that is equal to the threshold is excluded. The threshold is typically either the median or the mean. A subsequence is defined as the longest non-empty fragment consisting of identical elements. For instance, in the sequence of 19 elements '1110000011110000111', there are five subsequences, three of which are ones and two of which are zeros. The Wald-Wolfowitz statistic is determined by counting the number of such subsequences. If the number of subsequences differs significantly from the expected number, it is possible to reject the statistical independence of the sequence elements.

## *Repetition Distance*

The tendency of people to avoid repetition in the task of generating random sequences can be used to create a measure of deviation from randomness. For example, consider the sequence "2, 3, 7, 8, 8, 7, 2, 3, 2". The answer "2" is repeated at the sixth position and then at the second position. "3" on the sixth, "7" on the third, and "8" on the first. All this can be presented in the form of a table. The theoretical repetition distance can be approximated in the form of a geometric distribution:

$$frequency(s)_{expected} = (1-p)^{s-1} \times p \times (n-1)$$

where s is the number of steps before repetition (lag), p is the probability of choosing a number, n is the length of the sequence.

Quantitative measures of repetition can be obtained as measures of the lags between repetitions (repetition distance): mean lag, median lag, modal lag, and the like.

## *Bartels rank test of randomness (RVN)*

This is the rank version of von Neumann's Ratio Test for Randomness. Statistics show how close consecutive values are to each other:

$$RVN = \frac{\sum_{i=1}^{n-1}(R_i - R_{i+1})^2}{\sum_{i=1}^{n}(R_i - (n+1)/2)^2}$$, where $R_i$ = rank($X_i$), i = 1, ..., n - ranks of elements $X_i$ of a sequence of n numbers.

## *Mann-Kendall Trend Test (t)*

For each $X_i$, we compute the number $n_i$ of elements $X_j$ that precede it (i>j) such that rank($X_i$) > rank($X_j$). The statistic is equal to

$$t = \sum_i n_i .$$

It can be used to find subsequences where the trend is most pronounced.

These metrics measure differentMann-Kendall Trend Test deviations from randomness and are not completely interchangeable, although they are statistically related. It is important to note that experimental studies have shown that when generating random numbers, the R-statistic is sensitive to the size of the dictionary, i.e., the number of alternatives. Additionally, measures like A and RNG depend significantly on the required speed of responses and are therefore statistically dependent on statistics like NSQ, TPI, and PL.

To illustrate the relationship between the aforementioned indicators, a correlation matrix has been presented in Table 1. This matrix was obtained through a computer simulation of random sequences comprising of digits ranging from 0 to 9, using the R computing software environment (R Core Team, 2016) to perform the simulation and indicator calculations. This matrix was obtained through a computer simulation of random sequences comprising of digits ranging from 0 to 9, using the R computing software environment (R Core Team, 2016) to perform the simulation and indicator calculations.

*Table 1.* Correlations between measures of deviation from randomness were analyzed for 10,000 random sequences of 1,000 elements from the interval of 0-9.

|        | R     | Coupon | RNG   | NSQ   | A     | CS1   | CS2   | TPI   | RUNS  | MeanRG | RVN  | CSS  | Signs | t     |
|--------|-------|--------|-------|-------|-------|-------|-------|-------|-------|--------|------|------|-------|-------|
| Coupon | 0.01  |        |       |       |       |       |       |       |       |        |      |      |       |       |
| RNG    | **0.44** | -0.01 |       |       |       |       |       |       |       |        |      |      |       |       |
| NSQ    | **0.04** | -0.02 | 0.09  |       |       |       |       |       |       |        |      |      |       |       |
| A      | 0.00  | -0.01  | -0.01 | -0.01 |       |       |       |       |       |        |      |      |       |       |
| CS1    | 0.01  | 0.00   | 0.00  | -0.01 | **0.84** |       |       |       |       |        |      |      |       |       |
| CS2    | 0.00  | 0.01   | 0.01  | 0.00  | **-0.17** | **-0.13** |       |       |       |        |      |      |       |       |
| TPI    | -0.02 | 0.00   | 0.00  | 0.00  | **-0.22** | **-0.16** | **-0.05** |       |       |        |      |      |       |       |
| RUNS   | 0.00  | -0.01  | 0.00  | -0.01 | **0.25** | **0.21** | **0.10** | **-0.49** |       |        |      |      |       |       |
| MeanRG | -0.03 | **-0.58** | -0.01 | 0.01  | 0.02  | 0.01  | -0.01 | 0.01  | 0.02  |        |      |      |       |       |
| RVN    | -0.01 | -0.01  | 0.00  | 0.01  | **-0.36** | **-0.29** | **-0.18** | **0.55** | 0.01  | 0.02   |      |      |       |       |
| CSS    | 0.00  | -0.01  | 0.00  | 0.00  | -0.01 | -0.01 | 0.00  | 0.01  | 0.00  | 0.01   | 0.01 |      |       |       |
| Signs  | 0.01  | -0.01  | 0.01  | -0.01 | 0.01  | 0.00  | 0.00  | -0.02 | **0.58** | 0.00   | -0.01 | 0.00 |       |       |
| t      | 0.00  | -0.01  | -0.01 | -0.01 | -0.01 | -0.01 | 0.00  | 0.00  | 0.00  | 0.01   | 0.01 | **0.71** | -0.01 |       |
| RWW    | -0.01 | -0.01  | 0.01  | 0.01  | **-0.39** | **-0.33** | **-0.16** | **0.40** | **-0.15** | 0.01   | **0.78** | 0.00 | 0.01  | 0.00  |

Notes: Significant correlations at the 0.0005 level (two-sided test) are shown in bold. The following indices were used: R (Redundancy index), Coupon (Coupon index), RNG (Random Number Generation Index), NSQ (Null-Score Quotient), A (Adjacency), CS1 and CS2 (Count scores), TPI (Turning Point Index), RUNS (Runs),

MeanRG (Mean Repetition Distance), RVN (Bartels Rank Test), CSS (Cox-Stewart Statistic), Signs (Sign Difference), t (Mann-Kendall Trend Test), and RWW (Wald–Wolfowitz Runs Test).

Table 1 shows that certain indicators carry similar information. For instance, the Adjacency (A) and counting score 1 (CS1) have a correlation coefficient of 0.84. Similarly, the Wald-Wolfitz statistic (RWW) and Bartels rank statistic (RVN) have a correlation coefficient of 0.78. Lastly, the Kendall rank statistic (t) and Cox-Stewart statistic (CSS) have a correlation coefficient of 0.71 and can be used interchangeably.

To compare the degree of deviation from randomness in different numerical sequence generation tasks, it is important to consider how the length of the sequence affects the values of the indicators. The average values of the indices were then calculated. In this study, we generated 10,000 sequences of random integers ranging from 0 to 9. The sequences varied in length from 10 to 1000 numbers, in increments of 50. The study found that several indices, including the Coupon, Adjacency, Turning Point Index, Runs, Bartels rank statistic, Wald-Wolfitz statistic, Sign Difference, were practically independent of sequence length, even for short sequences (n=10, 60, 110...). Please refer to Figure 1 for the behavior of the remaining indices. Only Counting score 1 (bottom panel) is presented, as it exhibits similar behavior to Counting score 2.

*Figure 1.* The behavior of deviation indices when altering the length of the generated random sequence.

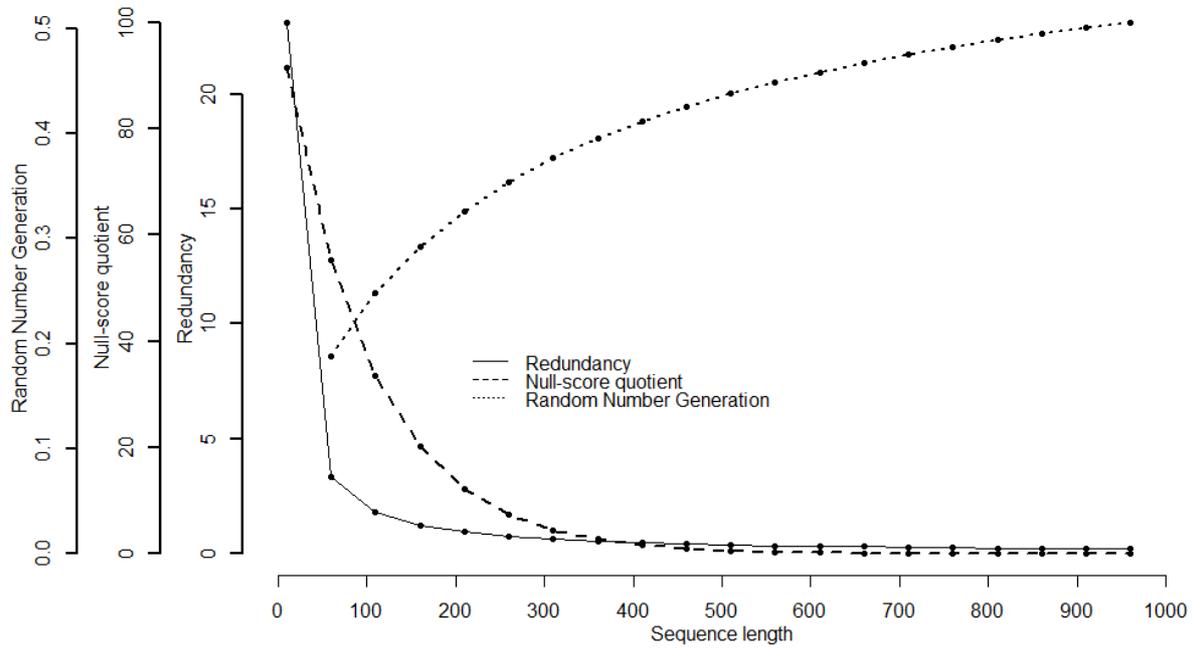

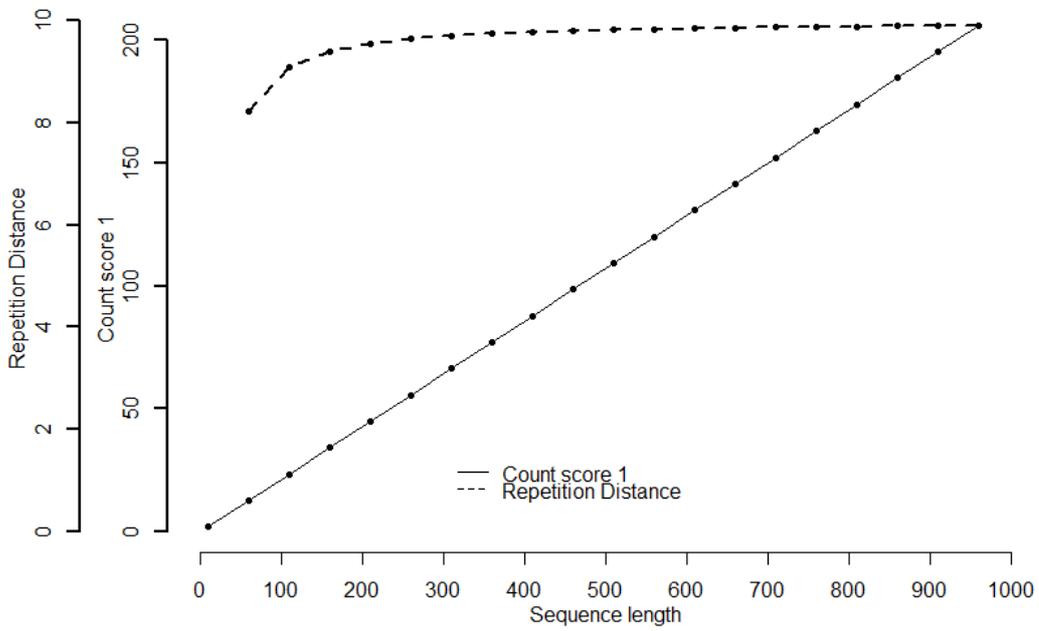

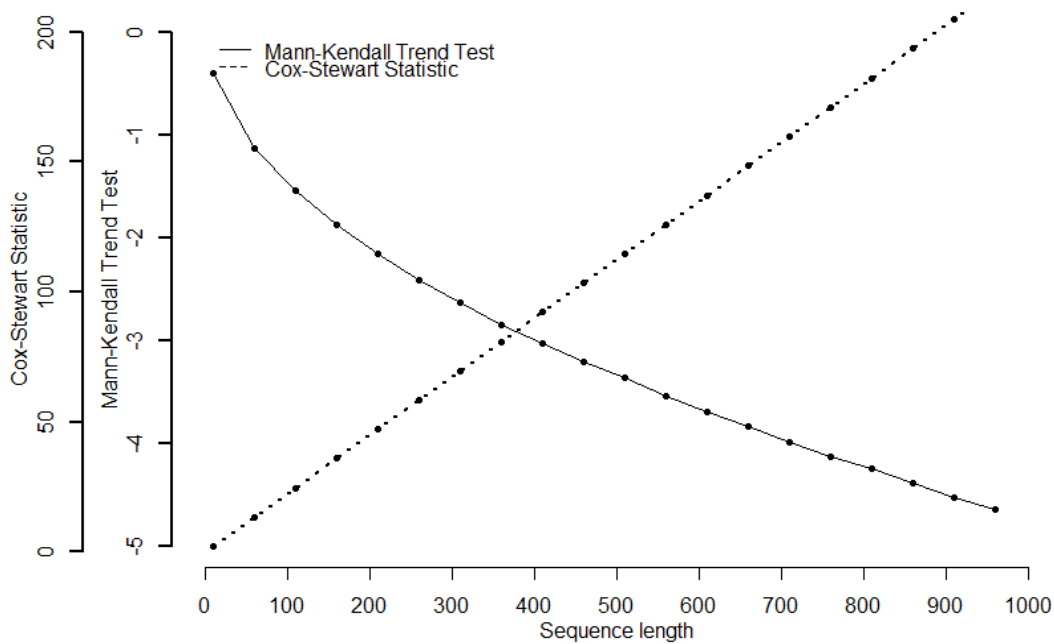

In order to analyze deviations from randomness, it is necessary to generate a set of random sequences with parameters (length, number of alternatives) similar to those used in the study. The average statistics of these random sequences can then be used as a basis for comparison.

941.

Wiegersma, S. Avoidance of repetition in produced sequences and the concept of randomness. *Perceptual & Motor Skills*. 1986, 62(1), 163-168.

Williams, M. A., Moss, S. A., Bradshaw, J. L., Rinehart, N. J. Random Number Generation in Autism. *Journal of Autism and Developmental Disorders*, 2002, 32(1), 43-47.